\documentclass[aps,prl,twocolumn,showpacs,groupedaddress,superscriptaddress]{revtex4-1}  
\usepackage[T1]{fontenc}
\usepackage[english]{babel}    		
\usepackage{amssymb}
\usepackage{amsmath}
\usepackage{graphicx}				
\usepackage{floatrow}
\usepackage[caption=false]{subfig}
\usepackage{multirow}
\usepackage{array}
\usepackage{verbatim}
\usepackage[svgnames]{xcolor} 
\usepackage{tikz}
\usepackage{float}

\begin{document}
\hspace{5.2in} 

\title{Knotted optical vortices in exact solutions to Maxwell's equations}
\author{Albertus J.J.M. de Klerk}
\affiliation{
   Huygens Kamerlingh Onnes Laboratory,
   Leiden University,
   P.O. Box 9504, 2300 RA Leiden, The Netherlands
   }
\affiliation{
   Mathematical Insitute, 
   Leiden University, 
   P.O. Box 9512, 2300 RA Leiden, The Netherlands
   }
\author{Roland I. van der Veen}
\affiliation{
   Mathematical Insitute, 
   Leiden University, 
   P.O. Box 9512, 2300 RA Leiden, The Netherlands
   }
\author{Jan Willem Dalhuisen}
\affiliation{
   Huygens Kamerlingh Onnes Laboratory,
   Leiden University,
   P.O. Box 9504, 2300 RA Leiden, The Netherlands
   }
\author{Dirk Bouwmeester}
\affiliation{
   Huygens Kamerlingh Onnes Laboratory,
   Leiden University,
   P.O. Box 9504, 2300 RA Leiden, The Netherlands
   }
\affiliation{
   Department of Physics, 
   University of California, 
   Santa Barbara, CA 93106, USA
   }
\date{\today}

\begin{abstract} 
	We construct a family of exact solutions to Maxwell's equations 
	in which the points of zero intensity form knotted lines topologically equivalent to a given but arbitrary algebraic link.
	These lines of zero intensity, more commonly referred to as optical vortices, and their topology 
	are preserved as time evolves and the fields have finite energy. 
	To derive explicit expressions for these new electromagnetic fields that satisfy the nullness property, we make use of the Bateman
	variables for the Hopf field as well as complex polynomials in two variables whose zero sets give rise to algebraic links. 
	The class of algebraic links includes not only all torus knots and links thereof, but also more intricate cable knots. 
	While the unknot has been considered before, the solutions presented here show that more general knotted structures 
	can also arise as optical vortices in exact solutions to Maxwell's equations.
\end{abstract}

\maketitle
\begin{figure}[htp]
	\sidesubfloat[]{
		\includegraphics[width=0.7\columnwidth]{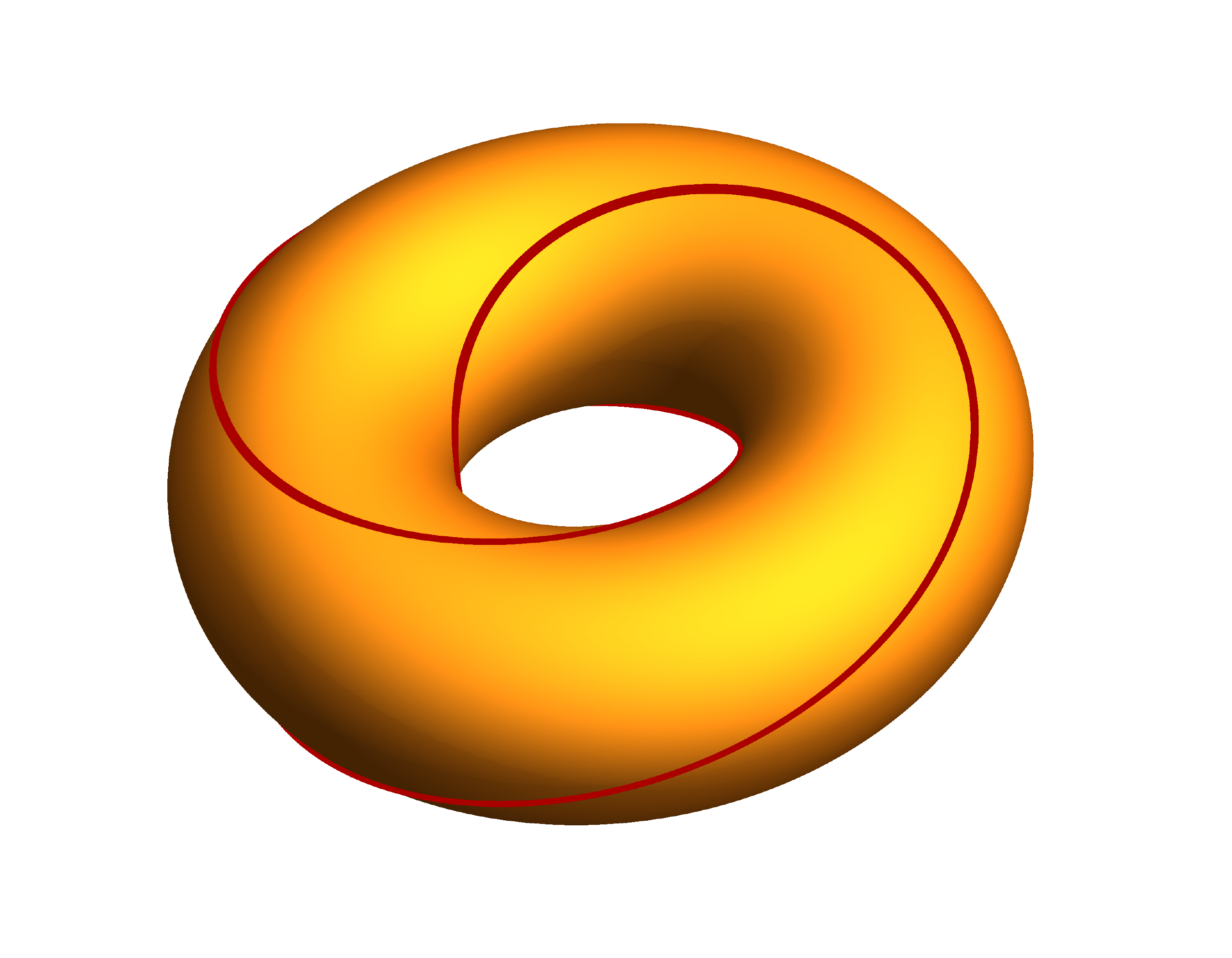}
		\label{fig:TorusKnot}
	}

	\sidesubfloat[]{
		\includegraphics[width=0.7\columnwidth]{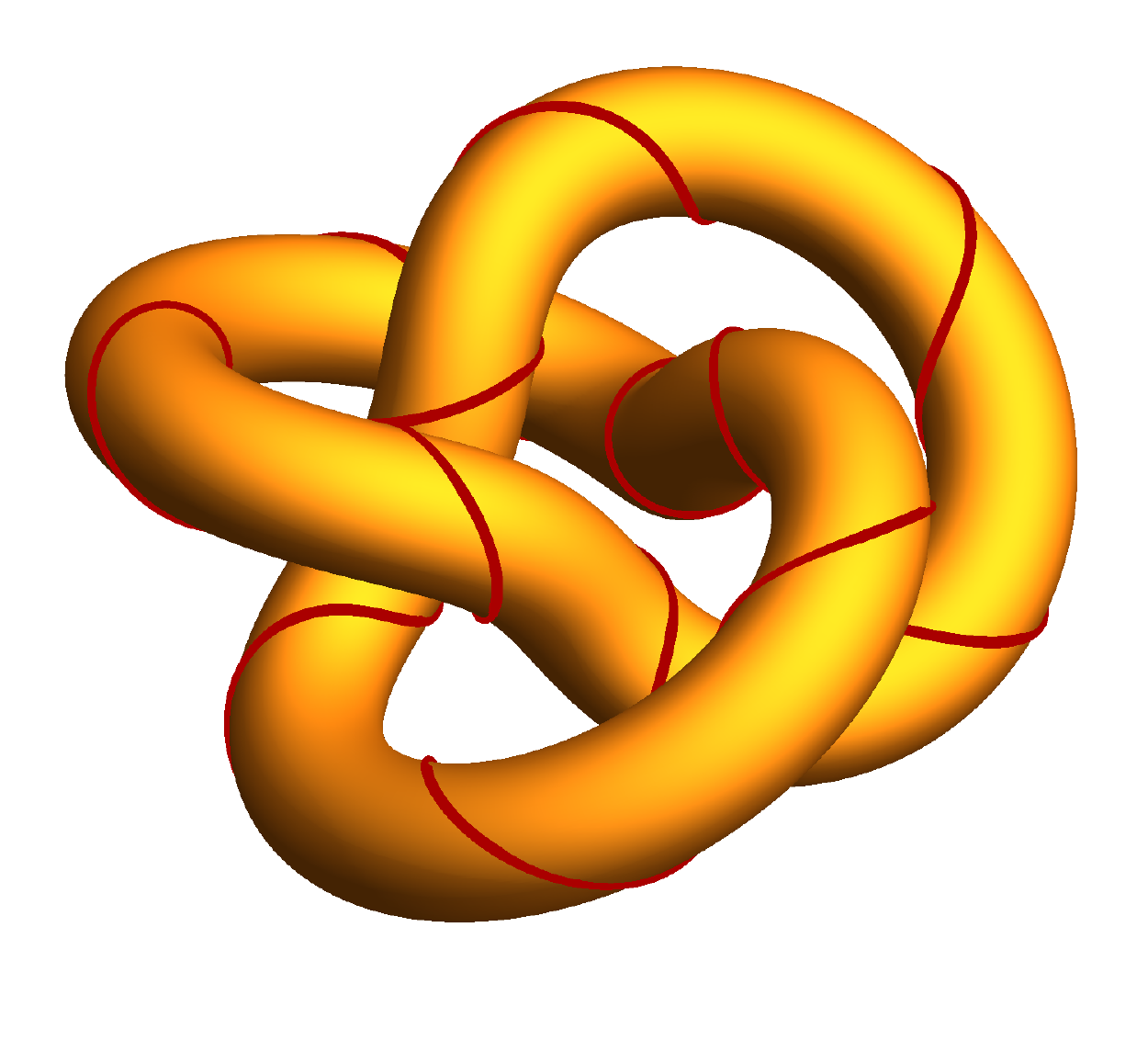}%
		\label{fig:CableKnot}
	}
	\caption{\textbf{A torus knot and a cable knot.}
	\textbf{a}, The red curve lying on the yellow torus is a (2,3) torus knot.
	\textbf{b}, The embedded yellow torus has a $(2,3)$ torus knot at its core. 
	The red curve is a cable knot with Newton pairs $(2,3)$ and $(3,2)$.}
	\label{fig:Torus and Cable knots}
\end{figure}

The discovery of the electromagnetic Hopf field by Ra\~nada \cite{Ranada1989,Ran1990,Ranada1992,Ranada1995} initiated further studies of knotted electromagnetic field lines \cite{Irvine2008,Besieris2009,Dalhuisen2012,Kedia2013,VanEnk2013,Arrayas,Hoyos2015}. This led to the formulation of a family of knotted electromagnetic fields encoding  torus knots by Kedia et al. \cite{Kedia2013}. These solutions were shown to be equivalent to elementary states arising in twistor theory \cite{Dalhuisen2012}, which allowed for a generalization of the knotted solutions to other massless field equations, in particular the linearized Einstein equations \cite{Thompson}. More recently Kedia et al. \cite{Kedia2016} have proposed a method capable of constructing divergence-free vector fields with knotted field lines more general than torus knots. However, a knotted structure can also be encoded in optical vortices, the lines of zero intensity of an electromagnetic field. These vortices and their topology have up to now primarily been studied in paraxial fields \cite{Berry2007,OHolleran2009,Leach2005,Dennis2009}. For example, Dennis et al. \cite{Dennis2010} employed the paraxial wave approximation to derive and produce knotted optical vortices. In contrast, Bialynicki-Birula used the Bateman construction \cite{Bateman1914} to introduce optical vortices topologically equivalent to lines and circles in exact solutions to Maxwell's equations \cite{Bialynicki-Birula2004}. Here we extend the class of known optical vortices in exact solutions to Maxwell's equations to include all algebraic links \cite{Milnor1968,Eisenbud1985,Baez1994,Wall2004}, providing the first examples of topologically non-trivial vortices in this context. \\ 
\indent \textit{Algebraic links --}
We start with an exposition of those aspects of knot theory necessary to describe the topology of the family of optical vortices constructed in this paper \cite{Milnor1968,Eisenbud1985,Baez1994,Wall2004}. Let $h$ be a polynomial in two complex variables with vanishing constant term. Then the zero set of this polynomial in $\mathbb{C}^2$ intersected with the 3-sphere of radius $\epsilon > 0$, denoted by $\mathbb{S}^3_\epsilon$, is diffeomorphic to a disjoint union of circles provided that $\epsilon$ is sufficiently small. This intersection is called an algebraic link, and the topology of this link is essentially independent of $\epsilon$. 
Thus, in particular, an algebraic link is a one-dimensional submanifold of $\mathbb{S}^3_\epsilon$ and can therefore be regarded as a one-dimensional submanifold of $\mathbb{R}^3$ via stereographic projection. Every component of such an algebraic link corresponds uniquely to an irreducible factor of the polynomial by which it is induced. Therefore, if the polynomial inducing the algebraic link is irreducible, the corresponding link has one component and is called an algebraic knot. 
\\ \indent 
There exists a method due to Newton to solve $h(v,w) = 0$ for $w$ in terms of $v$ for any irreducible polynomial $h$ that vanishes at the origin. This method gives an explicit algorithm to obtain successive approximations to the exact solution of $h(v,w)=0$ for $w$ in terms of $v$ of the form 
	\begin{align}
		\label{eq:approximations}
		w_0		&= a_0v^{q_0/p_0} \nonumber \\
		w_1 		&= v^{q_0/p_0}( a_0 + a_1 v^{q_1/p_1}) \\
		\vdots \nonumber 
	\end{align} 
for some complex numbers $a_0,a_1,\dotsc$. The pairs of coprime numbers $(p_i,q_i)$ determining the exponents in the expansion are called the Newton pairs, of which a finite number determine the topology of the knot. Therefore, some approximation $w_k$ containing all the information relevant to the topology of the knot can be obtained in a finite number of steps. \\ 
\indent For example, consider the general case for one Newton pair $(p,q)$ with corresponding polynomial 
	\begin{equation}
		h(v,w) = \sqrt{2}^q v^{q} - \sqrt{2}^p w^{p} 
	\end{equation}
In this case we can take $\epsilon$ equal to one and we obtain a parametrization $(v,w)$ for the knot by plugging $v = e^{i \theta p}/\sqrt{2}$ into $w = \sqrt{2}^{q/p}v^{q/p}$. This curve lies on the standard torus and closes after going $p$ and $q$ times around the toroidal direction and poloidal direction respectively. Such a knot is called a $(p,q)$ torus knot of which an example is illustrated in Fig. \ref{fig:TorusKnot}. 

\floatsetup[figure]{style=plain,subcapbesideposition=top}

However, care should be taken when trying to obtain parametrizations of knots associated to more general Newton pairs in this way. This is due to the fact that plugging some $v$ of absolute value $\epsilon$ into an expression for $w$ consisting of multiple terms will in general result in $(v,w)$ not being a subset of $\mathbb{S}^3_\epsilon$. To resolve this issue, it is common to deform $\mathbb{S}^3_\epsilon$ into a union of two solid tori:
	\begin{equation}
		\label{eq:1}
		\{ (v,w) \in \mathbb{C}^2 : |v|=\epsilon, |w| \leq \delta \textrm{ or } |v|\leq \epsilon, |w| = \delta \} 
	\end{equation}
This deformation can be performed in such a way that the zero set of the polynomial $h$ corresponding to the knot under consideration intersects this set in only one of the two tori. Furthermore, it can be shown that this intersection is topologically equivalent to the knot obtained by intersecting the zero set of $h$ with $\mathbb{S}^3_\epsilon$ provided that $\epsilon$ is sufficiently small.
\\ \indent Now consider a more general knot described by two Newton pairs $(p_0,q_0)$ and $(p_1,q_1)$. In this case we can plug $v  = \epsilon e^{i \theta p_1 p_2}$ into the expression for $w_1$ to obtain a parametrization for the knot corresponding to the given Newton pairs due to the discussion in the previous paragraph. The additional term in $w_1$ with respect to $w_0$ can be interpreted as perturbing the $(p_0,q_0)$ torus knot described by $(v,w_0)$. The resulting knot can then be described as a curve in a tubular neighborhood of the $(p_0,q_0)$ torus knot, which is an embedded solid torus with the torus knot at its core. To give a precise description of the knot itself, imagine a $(p_1,q_1)$ torus knot on a torus $T$ and embed this torus in such a way that it is a tubular neighborhood of the $(p_0,q_0)$ torus knot and such that the image of a curve with all its tangent vectors in the toroidal direction on $T$ is unlinked with the core of the embedding of $T$. Then the knot corresponding to the given Newton pairs is obtained by applying the same embedding to the $(p_1,q_1)$ torus knot on $T$. This process is demonstrated in Fig. \ref{fig:Torus and Cable knots}, where we start with a torus knot in Fig. \ref{fig:TorusKnot} and illustrate its tubular neighborhood and a cable knot corresponding to the additional Newton pair $(3,2)$ in Fig. \ref{fig:CableKnot}. This procedure can be repeated by embedding a torus such that its core is a cable knot if there are additional Newton pairs. Alternatively, one can start from some fixed Newton pairs and construct a corresponding polynomial in two complex variables as described in \cite{Eisenbud1985}. 

\textit{Algebraic links in electromagnetism --}
Now we will go on to show how algebraic links can be implemented as optical vortices in electromagnetic fields. To do so, we will describe an electromagnetic field with its Riemann-Silberstein vector
	\begin{equation} 
		\mathbf{F} = \mathbf{E} + i \mathbf{B} 
	\end{equation}
Provided that an electromagnetic field is null, i.e. satisfies $\mathbf{F} \cdot \mathbf{F} = 0$ which means that the fundamental invariants of the electromagnetic field $\mathbf{E} \cdot \mathbf{B}=0$ and $|\mathbf{E}|^2  - |\mathbf{B}|^2$ are zero, there exist smooth complex-valued functions $\alpha,\beta$ on Minkowsi space satisfying 
	\begin{equation}
		\nabla \alpha \times \nabla \beta = \pm i (\partial_t \alpha \nabla \beta - \partial_t \beta \nabla \alpha)
	\end{equation}
for which the field can be written as $\mathbf{F} = \nabla \alpha \times \nabla \beta$, as shown by Hogan \cite{Hogan1984}. Given such functions $\alpha$ and $\beta$, which we will refer to as Bateman variables, as well as arbitrary smooth maps $f$ and $g$ from $\mathbb{C}^2$ to $\mathbb{C}$, it holds that $f(\alpha,\beta)$ and $g(\alpha,\beta)$ are also Bateman variables \cite{Bateman1914}. The Bateman variables for the Hopf field seem to have originated from work by Robinson and Trautman \cite{Trautman} and  Bialinicki-Birula  \cite{Bialynicki-Birula2004}, which was expanded upon by Besieris and Sharaawi \cite{Besieris2009}, Van Enk \cite{VanEnk2013}, and Kedia et al. \cite{Kedia2013}. Here we will use a specific choice of Bateman variables for the Hopf field that is given in \cite{Kedia2013}:
	\begin{equation}
		\alpha = \frac{r^2-t^2-1+2iz}{r^2-(t-i)^2} \hspace{0.5cm} \textrm{ and } \hspace{0.5cm} \beta = \frac{2(x-iy)}{r^2-(t-i)^2}
		\label{eq:BatemanVariablesHopfField}
	\end{equation}
How the Hopf field and its Bateman variables can be obtained from a solution of the scalar wave equation is described in the supplemental information. By multiplying $\alpha$ and $\beta$ with $\epsilon/\sqrt{2}$ for any $\epsilon >0$, we obtain Bateman variables $\alpha_\epsilon$ and $\beta_\epsilon$ for a scaled version of the Hopf field $\mathbf{F}_H$. Then the map $(\alpha_\epsilon,\beta_\epsilon)$, restricted to a fixed time, can be interpreted as a map from $\mathbb{R}^3$ to $\mathbb{S}_\epsilon^3$ because $|\alpha_\epsilon|^2 + |\beta_\epsilon|^2 = \epsilon^2$ holds at any fixed time. To obtain an electromagnetic field with an optical vortex that is topologically equivalent to a given algebraic link $L$, we take a corresponding polynomial $h$ and $\epsilon >0 $ and choose 
\begin{equation}
	f(v,w) = \int h(v,w) dv \hspace{0.5cm} \textrm{ and } \hspace{0.5cm} g(v,w) = w 
\end{equation}
Then, as we will now show, the null field given by 
	\begin{align}
		\mathbf{F}_L 		&= \nabla f(\alpha_\epsilon,\beta_\epsilon) \times \nabla g(\alpha_\epsilon,\beta_\epsilon) \nonumber \\
						&= h(\alpha_\epsilon,\beta_\epsilon) \nabla \alpha_\epsilon \times \nabla \beta_\epsilon
	\end{align}
has optical vortices topologically equivalent to $L$ at every time. This construction is possible by virtue of the fact that the intensity of the Hopf field is never zero and the fact that Bateman variables of the Hopf field can be written as done in equation \ref{eq:BatemanVariablesHopfField}, which at $t=0$ describe the inverse stereographic projection. \\
\indent
First, we note that since the intensity of the Hopf field is nowhere zero, the optical vortices of $\mathbf{F}_L$ are determined completely by the zero set of $h(\alpha_\epsilon,\beta_\epsilon)$. Secondly, the fact that restricted to $t=0$, the map $(\alpha_\epsilon, \beta_\epsilon)$ describes the inverse stereographic projection implies that the optical vortices of $\mathbf{F}_L$ at $t=0$ are topologically equivalent to the zero set of $h$ in $\mathbb{S}_\epsilon^3$, provided that the latter does not contain $(1,0)$. If it does, we can choose another polynomial giving rise to the same link for which this is not the case. Thirdly, it turns out that $(\alpha_\epsilon,\beta_\epsilon)$ restricted to any fixed time $t_*$ has rank three, which means that it is a local diffeomorphism from $\{t_*\} \times \mathbb{R}^3$ to $\mathbb{S}^3_\epsilon$. Thus, the zero set of $h(\alpha_\epsilon,\beta_\epsilon)$ at any fixed time is diffeomorphic to a disjoint union of circles and hence a link. Finally, it should be noted that the restriction of $h$ to $\mathbb{S}^3_\epsilon$ has rank two which, together with the fact that $(\alpha_\epsilon,\beta_\epsilon)$ restricted to a fixed time $t_*$ is a local diffeomorphism, implies that the zero set of $h(\alpha_\epsilon,\beta_\epsilon)$ in Minkowski space is a 2-dimensional manifold. This implies that the zero set of $h(\alpha_\epsilon,\beta_\epsilon)$ at any fixed time is topologically equivalent to the zero set of $h(\alpha_\epsilon,\beta_\epsilon)$ at $t=0$, which, in turn, is equivalent to $L$ as mentioned above. 
We note that the fields under consideration here fall into a class for which it is known that the structure is transported by a rescaled nowhere zero version of the Poynting vector field \cite{Newcomb1958,Irvine2010}. This observation leads to the same conclusion as our proof above. 

\begin{figure}[htp]
	\sidesubfloat[]{
		\includegraphics[width=0.7\columnwidth]{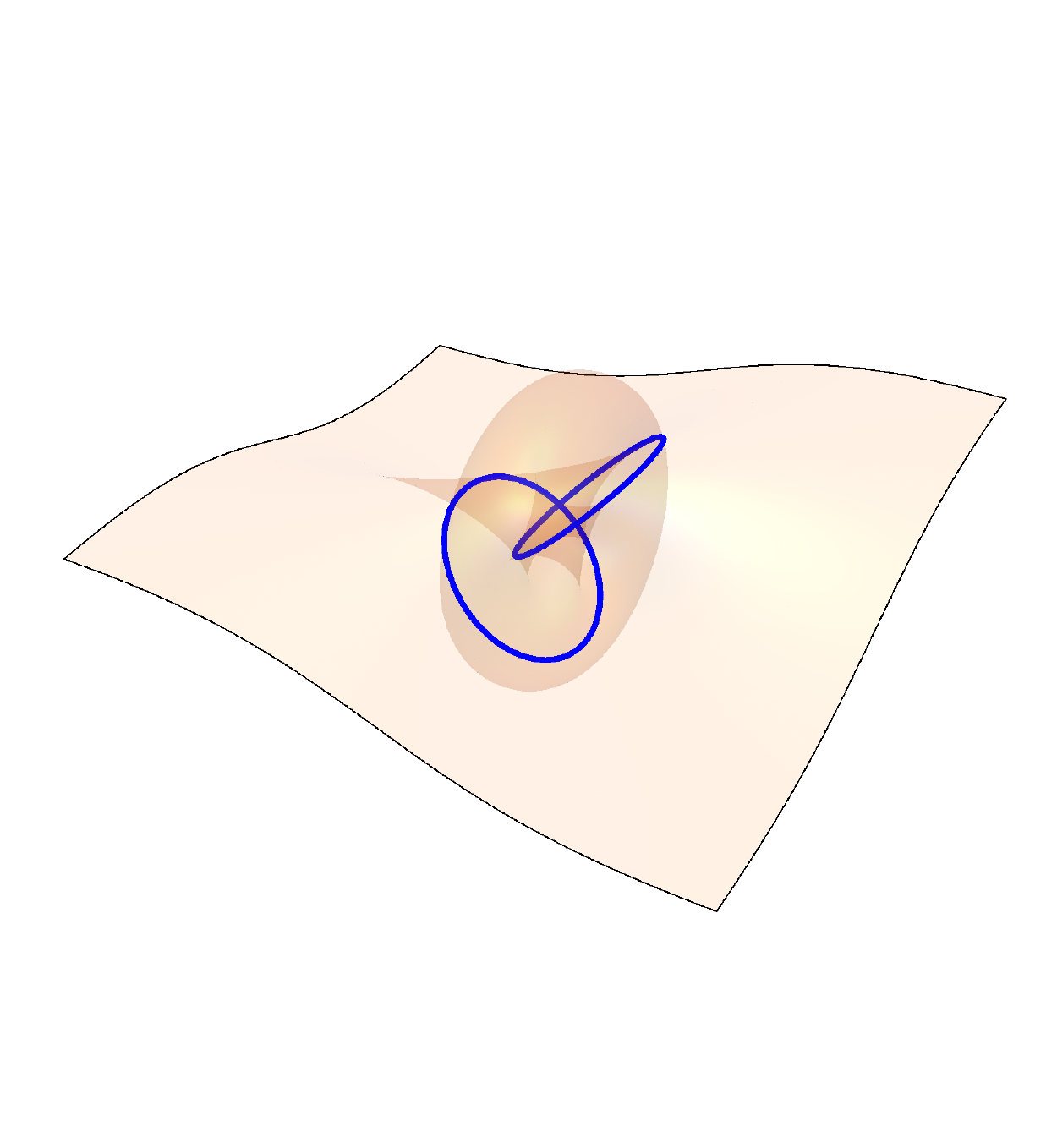}
	}

	\sidesubfloat[]{
		\includegraphics[width=0.7\columnwidth]{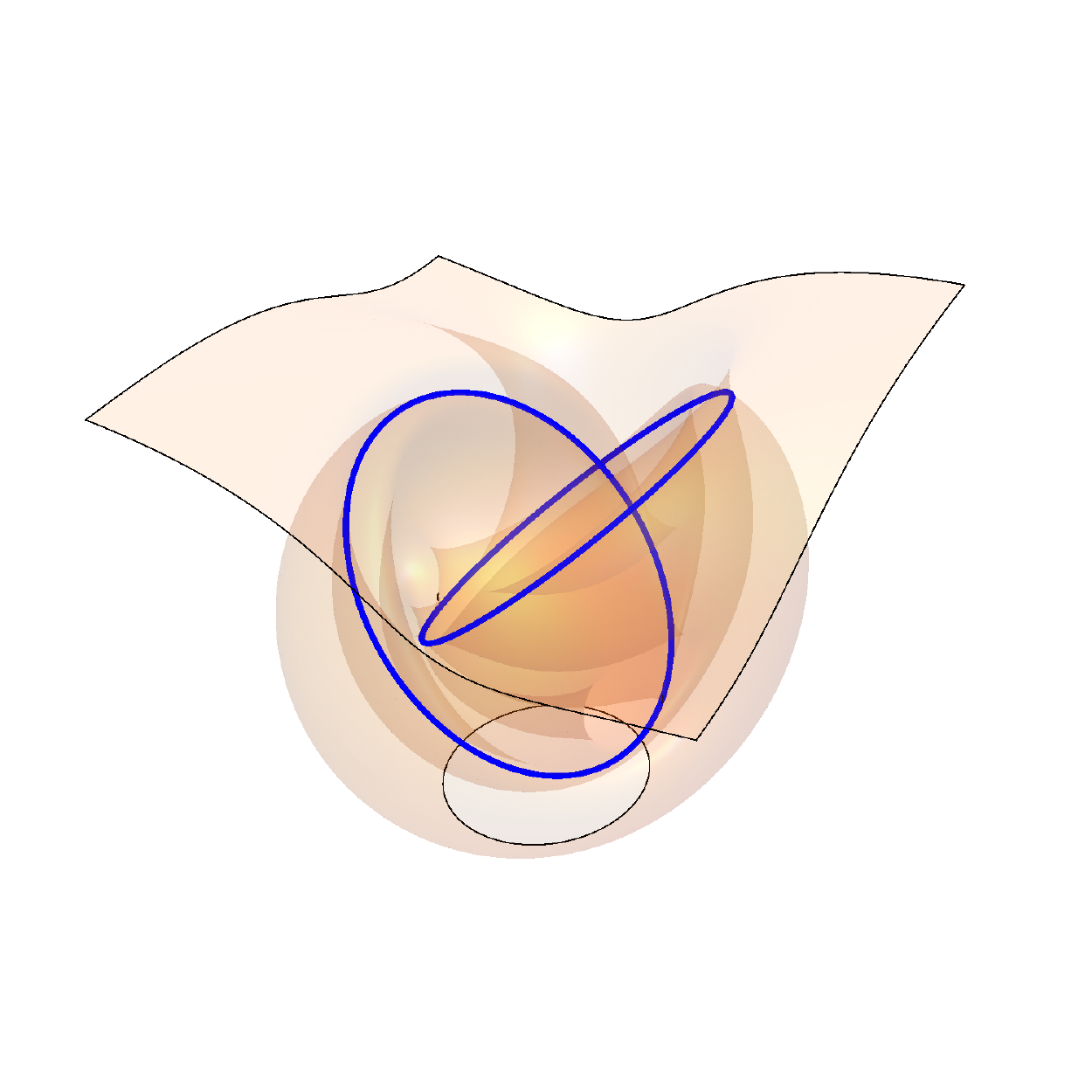}
	}
	\caption{\textbf{The Hopf link as an optical vortex.} 
	These plots are created by taking the intersection of the surfaces where the real and imaginary parts of $h_1$ are zero at $t=0$ (\textbf{a}) and $t=3$ (\textbf{b}). 
	The aforementioned surfaces are included for clarity and are denoted by the transparent orange surfaces.} 
	\label{fig:HopfLink}
\end{figure}

We will now examine some general properties of the class of electromagnetic fields we have constructed. First, we note that the energy density of an electromagnetic field $\mathbf{F}_L$ with corresponding polynomial $h$ and $\epsilon$ is given by 
	\begin{equation}
		\mathbf{F}_L \cdot \tilde{\mathbf{F}}_L = |h(\alpha_\epsilon,\beta_\epsilon)|^2 \textrm{ } \mathbf{F}_H \cdot \tilde{\mathbf{F}}_H 
	\end{equation} 
where $\tilde{\mathbf{F}}$ denotes the complex conjugate of $\mathbf{F}$. Since the energy of the Hopf field is finite and $|h(\alpha_\epsilon,\beta_\epsilon)|^2$ is bounded from above, we can conclude that the energy of $\mathbf{F}_L$ is finite as well. Secondly, the Poynting vector is given by 
	\begin{align}
		\mathbf{S}_L 	&= \textrm{Re}[\mathbf{F}_L] \times \textrm{Im}[\mathbf{F}_L] \nonumber \\
					&= |h(\alpha_\epsilon,\beta_\epsilon)|^2 \textrm{ } \textrm{Re}[\mathbf{F}_H] \times \textrm{Im}[\mathbf{F}_H]
	\end{align} 
Thus, away from the zero set of the Poynting vector field which is topologically equivalent to the algebraic link $L$,  the integral curves of the Poynting vector have the structure of the Hopf fibration. Also, since the momentum of the Hopf field is finite, the same holds for the momentum of $\mathbf{F}_L$. 

\captionsetup[subfigure]{labelfont={color=RoyalBlue,bf},labelformat=empty}
\begin{figure*}
	\scalebox{1}{
	\centering
	\begin{tabular}{cccc}
		\sidesubfloat[)]{\includegraphics[width=0.2\columnwidth]{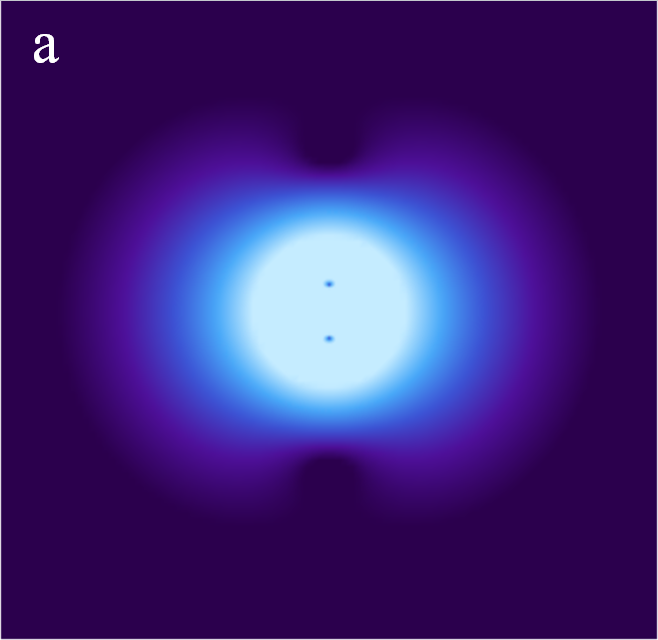}} & 
		\sidesubfloat[]{\includegraphics[width=0.2\columnwidth]{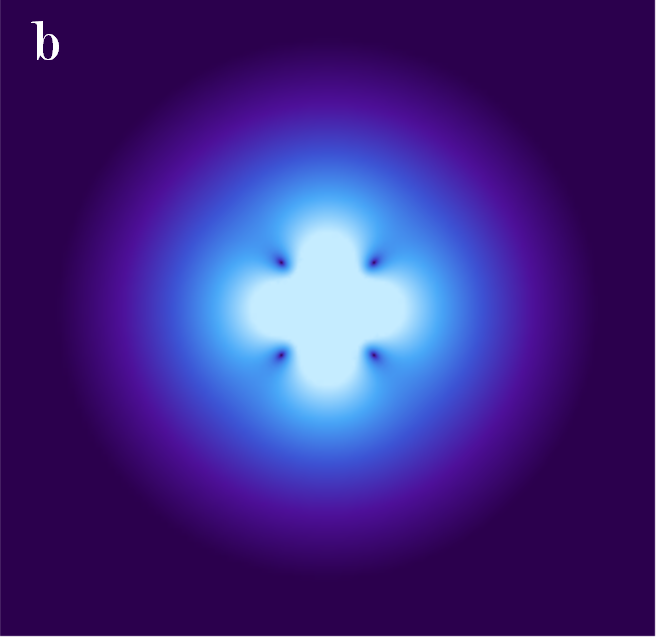}} & 
		\sidesubfloat[]{\includegraphics[width=0.2\columnwidth]{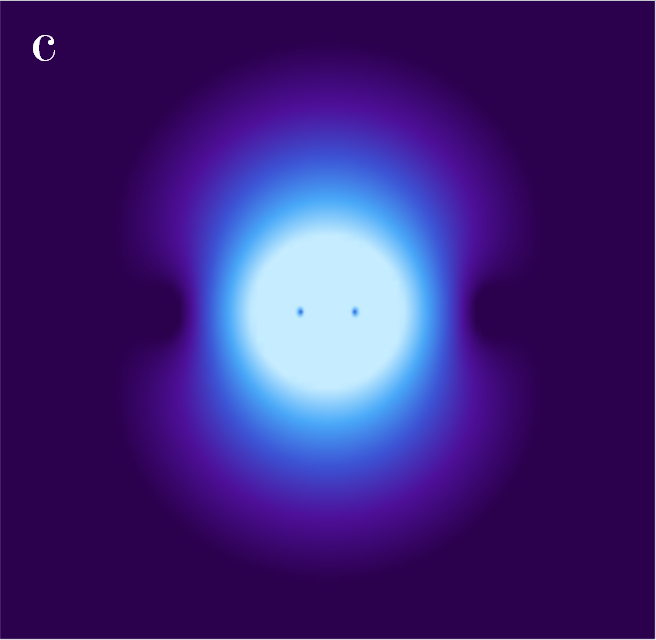}} & 
		\multirow{2}[-1]{*}{\sidesubfloat[]{\includegraphics[width=1.5cm,height=7.25cm,trim=0 0 0 0.9cm]{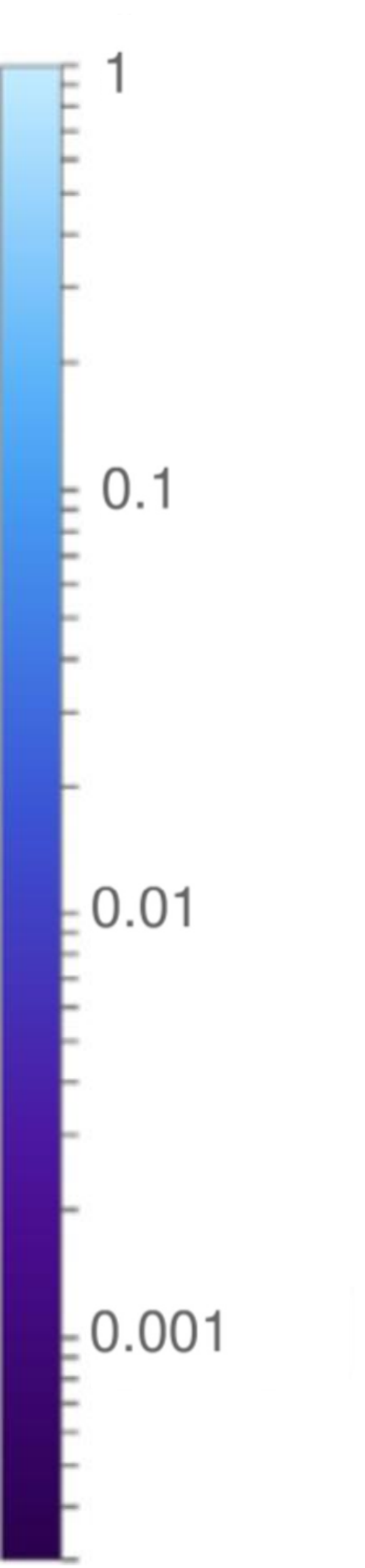}}} \\
		\sidesubfloat[]{\includegraphics[width=0.2\columnwidth]{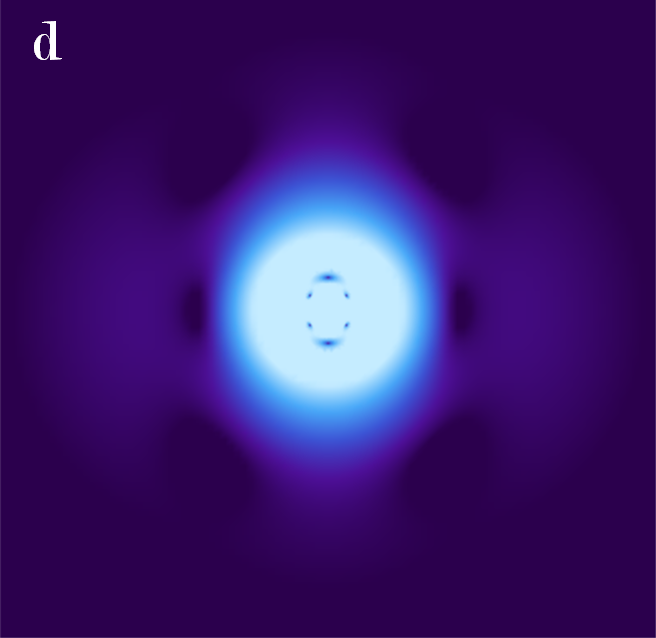}} & 
		\sidesubfloat[]{\includegraphics[width=0.2\columnwidth]{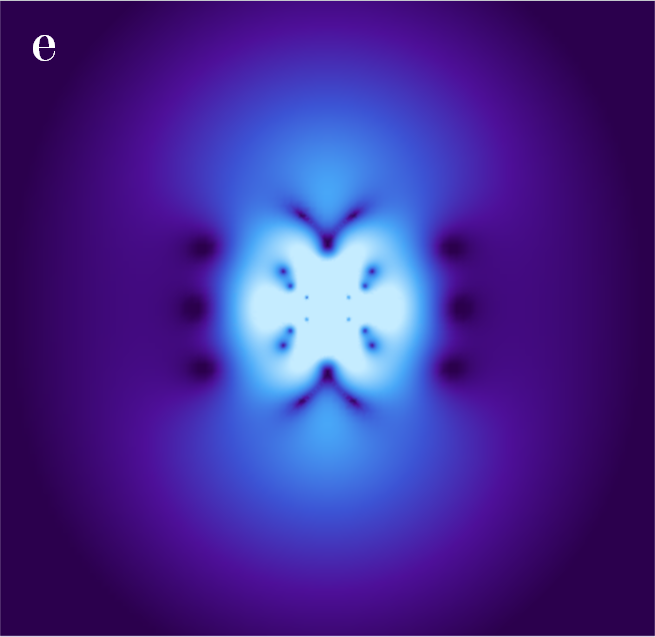}} & 
		\sidesubfloat[]{\includegraphics[width=0.2\columnwidth]{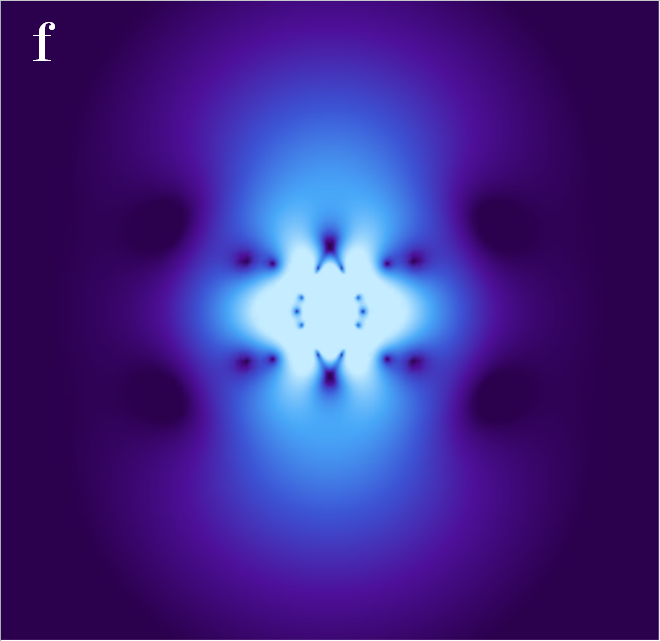}}\\
	\end{tabular}
	\caption{\textbf{Energy densities corresponding to $\mathbf{F}_1$ and $\mathbf{F}_2$.}
	\textbf{a} - \textbf{c},
	Logarithmic plot of the energy density of the field corresponding to equation (\ref{eq:Hopf Link}) in the xy, xz, and yz planes respectively. 
	\textbf{d} - \textbf{f},
	Logarithmic plot of the energy density of the field corresponding to equation (\ref{eq:Cable Knot}) in the xy, xz, and yz planes respectively.}
	\label{fig:Energy Densities}	}
\end{figure*}

The vector potentials for the electric and magnetic fields can be described by a complex vector 
	\begin{equation}
		\mathbf{V} = \mathbf{C} + i \mathbf{A} 
	\end{equation}
where $\mathbf{C}$ and $\mathbf{A}$ are given by the real and imaginary parts of $ f(\alpha,\beta) \nabla \beta$ respectively. Since the fields under consideration are null, the electric and magnetic helicities, which can be computed from the vector potentials, are conserved quantities \cite{Hoyos2015}. 

Now we will illustrate some of the general properties for the  Hopf link and a cable knot. The polynomial corresponding to the Hopf link is given by 
	\begin{equation}
		\label{eq:Hopf Link}
		h_1(v,w) = v^2  + w^2
	\end{equation}
The polynomial corresponding to a cable knot with Newton pairs (2,3) and (3,2) is given by 
	\begin{align}
		\label{eq:Cable Knot}
		h_2(v,w) 	&= w^6-3w^4v^3+3w^2v^6 \nonumber \\
		 		&-6w^2v^8-v^9-2v^{11}-v^{13}
	\end{align}
Then $\mathbf{F}_1 = h_1(\alpha,\beta) \nabla \alpha \times \nabla \beta$ and $\mathbf{F}_2 = h_2(\alpha,\beta) \nabla \alpha \times \nabla \beta$ are electromagnetic fields with optical vortices that are at any time topologically equivalent to the Hopf link and a cable knot with Newton pairs (2,3) and (3,2) respectively. The evolution of the vortex in $\mathbf{F}_1$ is illustrated in Fig. \ref{fig:HopfLink}. Furthermore, the energy densities at $t=0$ of both $\mathbf{F}_1$ and $\mathbf{F}_2$ are illustrated in Fig. \ref{fig:Energy Densities}. \\ 
\indent \textit{Summary --} We have derived analytic expressions for a new family of finite energy null electromagnetic fields with optical vortices that are topologically equivalent to algebraic links. Furthermore we provide a rigorous explanation for the emergence of these new topologically non-trivial structures. Since our solutions are based on the Bateman construction, it follows from a recent result by Goulart \cite{Goulart2016} that they are also exact solutions of non-linear electrodynamics. We expect that the implementation of knot theory to generate topologically non-trivial structures in electromagnetism introduced here will lead to generalizations in linearized gravity \cite{Thompson,Nichols2011}, Bose-Einstein condensate configurations \cite{Kawaguchi2008}, and plasma configurations \cite{Thompson2014,Kamchatnov1982}. \\
\indent \textit{Acknowledgements --} 
This work is supported by the Netherlands Organisation for Scientific Research (NWO) through NWO VICI grant no. 680-47-604 and Spinoza Award 2014, and by the National Science Foundation NSF PHY-120611



%

\section{Appendix: Supplementary Material} 

The Bateman variables for the electromagnetic Hopf field have been found and presented in literature without an explicit derivation. 
Here we show how the Bateman variables emerge naturally when the Hopf field is derived from a superpotential. 

Throughout this supplementary material, we will describe electromagnetic fields with differential 2-forms on Minkowski space.
In this formalism \cite{Baez1994App}, Maxwell's equations in free space are given by 
\begin{equation}
	d F = 0 \hspace{0.5cm} \textrm{ and } \hspace{0.5cm} d \star F = 0
\end{equation}
where $\star$ is the Hodge star operator induced by the metric. 
The Hodge star operator on Minkowski space satisfies $\star^2 F = - F$ for any 2-form $F$. 
Therefore, the eigenvalues of $\star$ are $\pm i$ which leads us to choose $F$ to be complex-valued. 
A real solution to Maxwell's equations can always be obtained from the complex-valued form due to linearity of Maxwell's equations. 
If $\star F = i F$ holds, we call the form self-dual and if $\star F = -iF$ holds, we call the form anti-self-dual. 
A complex-valued 2-form is a solution to Maxwell's equations if it is (anti-)self-dual and closed. 
Given smooth maps $\alpha,\beta: \mathcal{M} \to \mathbb{C}$, the 2-form $F = d \alpha \wedge d \beta$ is closed by construction but only (anti-)self-dual if 
\begin{equation}
	\nabla \alpha \times \nabla \beta = \pm i (\partial_t \alpha \nabla \beta - \partial_t \beta \nabla \alpha)
\end{equation}
Provided that $\alpha$ and $\beta$ satisfy this equation, one can easily show that $F = d f(\alpha,\beta) \wedge d g(\alpha,\beta)$ is also a solution to Maxwell's equations for arbitrary smooth maps $f,g: \mathbb{C}^2 \to \mathbb{C}$. This shows how the Bateman construction arises naturally from the formalism employed here.

Now we will show how a solution to Maxwell's equations can be constructed from a superpotential, i.e. a solution to the scalar wave equation $\Delta W = (d \delta  + \delta d ) W =0$. This construction, for which we base ourselves on chapter 9 of Synge's book on special relativity \cite{Synge1956}, can be formulated in the formalism of differential forms as follows. Let $W$ be a solution of the scalar wave equation, and let $K$ be a constant 2-form on Minkowski space, then 
\begin{equation}
	A = \star (dW \wedge K)
\end{equation}
is a potential. To see this, note that this 1-form satisfies the Lorentz gauge condition $\delta A =0$ which implies that it is a potential since its components satisfy the wave equation. In the same chapter, Synge notes that 
\begin{equation}
	W = (r^2-(t-i)^2)^{-1}
\end{equation}
is a solution of the wave equation without singularities. If we take $K$ to be given by 
\begin{equation}
	K = - dz \wedge dx - i \cdot dy \wedge dz - dx \wedge dt + i \cdot dy \wedge dt
\end{equation}
the construction described above gives a potential given by
\begin{align}
	A 	&= \frac{-2 (y+i x)}{\left(x^2+y^2+z^2-(t-i)^2 \right)^2}dz \nonumber \\
		&+ \frac{2 (y+i x)}{\left(x^2+y^2+z^2-(t-i)^2 \right)^2}dt \nonumber \\
		&+ \frac{-2 i t+2 i z-2}{\left(x^2+y^2+z^2-(t-i)^2 \right)^2}dx \nonumber \\ 
		&+ \frac{-2 t+2 z+2 i}{\left(x^2+y^2+z^2-(t-i)^2 \right)^2}dy 
	\label{eq:potential Hopf}
\end{align}
The field $F = dA$ corresponding to this potential is then a self-dual 2-form for which the corresponding electric and magnetic fields are that of the Hopf field. 
Note that for an electromagnetic field of the form $F = d \tilde \alpha \wedge d \tilde \beta$, potentials are given by $A = \tilde \alpha d \tilde \beta$ and $A' = - \tilde \beta d \tilde \alpha$. 
Also, in equation (\ref{eq:potential Hopf}) there are four functions that serve as components, but two of them are related by a minus sign and the other two by a factor $i$. 
However, these two functions cannot be Bateman variables since the denominator of the product of one of these functions with the derivative of the other will be of higher order than the denominator of $A$. 
Therefore it is natural to consider the same functions without the square in the denominator. These maps are indeed Bateman variables for the Hopf field and are given by 
\begin{align}
	\tilde{\alpha}(t,x,y,z) 	&= \frac{2i-2t+2z}{r^2-(t-i)^2}  \\ 
	\tilde{\beta}(t,x,y,z) 	&= \frac{2(ix+y)}{r^2-(t-i)^2}
\end{align}
To obtain the Bateman variables used in the main article from the Bateman variables derived here, one has to take a factor $i$ from $\tilde{\beta}$ to $\tilde{\alpha}$ and add one to $i \tilde{\alpha}$. 
This explicit form is necessary in the main text as it allows us to interpret $(\alpha,\beta)$ as a map into the 3-sphere.

\end{document}